\documentclass[apj]{emulateapj}
\usepackage{natbib}
\usepackage{color}
\usepackage{amssymb}
\usepackage{amsmath}
\usepackage{graphicx}

\begin{document}

\title{Type II Migration of Planets on Eccentric Orbits}

\author{Althea V. Moorhead and Eric B. Ford}
\affil{Astronomy Department, University of Florida,
    Gainesville, FL 32611}

\begin{abstract}
The observed extrasolar planets possess both large masses (with a median $M \sin i$ of 1.65 $M_J$) and a wide range in orbital eccentricity ($0 < e < 0.94$). As planets are thought to form in circumstellar disks, one important question in planet formation is determining whether, and to what degree, a gaseous disk affects an eccentric planet's orbit.  Recent studies have probed the interaction between a disk and a terrestrial planet on an eccentric orbit, and the interaction between a disk and a gas giant on a nearly circular orbit, but little is known about the interaction between a disk and an eccentric gas giant.  Such a scenario could arise due to scattering while the disk is still present, or perhaps through planet formation via gravitational instability. We fill this gap with simulations of eccentric, massive (gap-forming) planets in disks using the hydrodynamical code FARGO.  Although the long-term orbital evolution of the planet depends on disk properties, including the boundary conditions used, the disk always acts to damp eccentricity when the planet is released into the disk.  This eccentricity damping takes place on a timescale of 40 years, 15 times faster than the migration timescale.
\end{abstract}


\section{Introduction}

The detection of over 200 extrasolar planets has overturned our understanding of what constitutes a typical planetary system.  In contrast to the solar system, the extrasolar planetary systems generally contain giant planets with small semi-major axes (0.017 AU $\le a \le$ 6 AU) and sizable eccentricities ($0 \le e \le 0.94$) \citep[e.g.,][]{2006ApJ...646..505B}.

The small semi-major axes of giant planets suggest that the many exoplanets did not form in situ. Rather, planets are assumed to form exterior to the ``snow-line," or ice condensation radius, and to subsequently migrate inward through interactions with a circumstellar disk \citep[e.g.,][]{2004ApJ...604..388I}.  This migration has two limiting cases in a laminar disk.  In Type I migration, a planet lacks sufficient mass to clear a gap in the disk material and is driven inward, and its eccentricity damped, by a density wake in the disk \citep{2000MNRAS.315..823P, 1993ApJ...419..166A, 2000MNRAS.315..823P, 2007A&A...473..329C}. In Type II migration, massive planets clear a gap in an annulus surrounding their orbits.  Analytic treatments of Type II migration generally consider the planet to be locked to the the viscous evolution of the disk \citep[e.g.,][]{1997Icar..126..261W}, although recent work by \cite{2008arXiv0807.0625E} shows that the planet is not as strongly coupled to the disk viscosity as predicted.  He claims that the inverse relationship between migration rate and planet mass, combined with a weak dependence on disk viscosity, indicates that Type II migration should be thought of as an angular momentum exchange between disk and planet, rather than the outer and inner disk using the planet as a simple relay station.  

The tendency of the planet to migrate inwards is well established, provided the planet does not lie exterior to the majority of the disk \citep[e.g.][]{1986ApJ...309..846L, 1997Icar..126..261W, 2008arXiv0807.0625E}.  The behavior of orbital eccentricity under the influence of a disk, on the other hand, is not well understood and oft debated.  While the aforementioned papers find that the disk damps eccentricity, some recent papers argue that the disk can, in fact, excite eccentricity \citep{2003ApJ...585.1024G, 2003ApJ...587..398O, 2006ApJ...652.1698D}.  However, previous studies started the planet on a circular or nearly circular orbit.  Most known exoplanets have substantial orbital eccentricity, and planets in multiple-planet systems may attain significant eccentricities due to planet-planet interactions while Type II migration is still underway.  Therefore, we turn our attention to Type II migration in the eccentric regime.

Previously, \cite{2008Icar..193..475M} performed calculations of Type II migration in the non-negligible eccentricity regime and concluded that orbital eccentricities up to values of 0.5 tend to be damped by the disk.  In this paper we perform hydrodynamical simulations of a giant planet on an eccentric orbit.  We describe our methods in \S\ref{sec:methods}.  In \S\ref{sec:diskprop}, we perform several tests to establish the sensitivity of our results to simulation choices, such as resolution, and disk properties, such as viscosity.  We present results showing that eccentricity is damped for all choices of disk properties in \S\ref{sec:orbit}, and discuss the implications of these results in \S\ref{sec:conclude}.

\section{Methods} 
\label{sec:methods}

All simulations presented in this paper show the orbital evolution of a Jupiter-mass planet in a minimum mass solar nebula.  To perform our hydrodynamic simulations, we use the FARGO code of \cite{2000A&AS..141..165M}, which is a 2D polar grid code tailored to planet-disk systems.  

All simulations are performed with the standard public release of FARGO\footnote{http://fargo.in2p3.fr/}, in which the disk is non-self-gravitating and isothermal.  The hydrodynamics are largely calculated in a ``standard'' manner \citep[i.e., consistent with][]{1992ApJS...80..753S}.  However, after calculation of the radial transport and the residual azimuthal transport, FARGO separates the disk into rings and calculates the azimuthal transport in the comoving frame of each ring, before recombining the rings in preparation for the next time step.  This allows the code to take time steps larger than the Courant-Friedrichs-Lewy condition, which requires that disk material cannot move further than one grid cell per time step.  A more detailed description of FARGO is given by \cite{2000A&AS..141..165M}, and an excellent synopsis is given by \cite{2008arXiv0807.0625E}.

Where sensible, we choose disk parameters equal to those in the \cite{2006MNRAS.370..529D} comparison project.  However, we must deviate from these choices in many cases where either: [1] the simulation of an eccentric planet requires it, or [2] we wish to investigate the possible effects a disk parameter has on the long term evolution of the planet's orbit.  As an example, we generally place our disk boundaries at 0.2 and 5.0 AU, rather than 0.4 and 2.5 AU, in order to accommodate the greater range in radial position of an eccentric planet.

This work takes an approach similar to that of \cite{2007A&A...473..329C}, in which the authors simulated the interaction between a disk and planets $10 M_{\oplus}$ in size with substantial initial eccentricity. As their work probes Type I migration, we focus on simulations with more massive planets ($> M_J$).  Initially, we place the planet in a smooth disk (i.e., no gap), but integrate the system with the planet's orbital elements held fixed until the disk settles and the planet clears a gap.

In all cases, our planet's mass is one Jupiter mass, and its initial semi-major axis is 1 AU.  In each run the planet is held on a fixed orbit for a period of time (100 years, unless stated otherwise), and then released so that we may measure its response to the disk.  This allows the disk to relax into a configuration that is consistent with the planet's eccentric orbit.  This initial period with fixed orbital elements is reasonable when the semi-major axis damping induced by the disk takes place on a timescale longer than that on which eccentricity is damped; eccentricity damping is calculated by \cite{2000MNRAS.315..823P} to be a factor of 100 faster than semi-major axis damping.

\section{Results}

We present two sets of results: First, we discuss the effects that varying disk properties has on planet-disk interactions and thus on the orbital evolution of the planet.  Second, we identify common behavior over different disk types and discuss the ramifications for planet migration.

\subsection{Disk Properties}
\label{sec:diskprop}

\subsubsection{Disk Viscosity}

\begin{figure}[t]
  \includegraphics[width=\columnwidth]{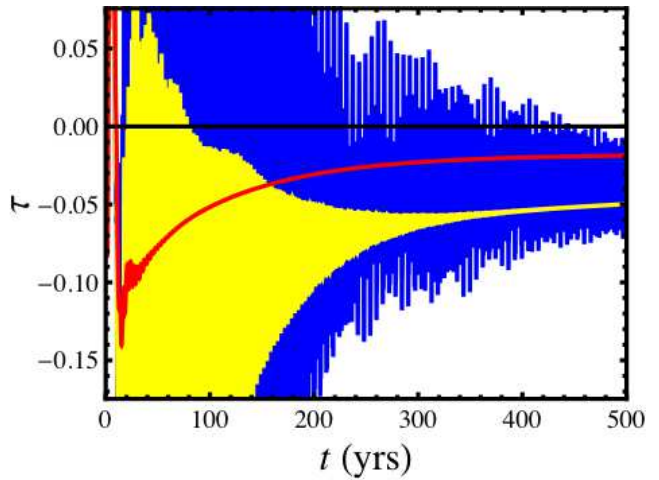}

  \caption[Disk torques]{Total torque exerted on the planet by the disk, for the cases of $\nu = 10^{-5}$ (yellow), $\alpha = 10^{-3}$ (blue), and $\alpha = 10^{-2}$ (red) when a Jupiter-mass planet moves on a fixed, circular orbit within the disk.  In future simulations we use $\alpha = 10^{-2}$ and let the disk relax for 100 years before releasing the planet.}
  \label{torques}
\end{figure}

First, we investigate the role of viscosity in planet-disk interactions using three choices: a constant viscosity of $\nu = 10^{-5}$ (in units where $a = G(M_{\ast} + M_P) = 1$, or $\nu = 4.5 \times 10^{7}~\mbox{m}^2/\mbox{s}$), and constant alpha-viscosities of $\alpha = 10^{-3}$ and $\alpha = 10^{-2}$.  The torques resulting from these three simulations are shown in Fig.~\ref{torques}.  In each case, the total torque exerted by the disk on the planet fluctuates with diminishing amplitude over time.  The timescale over which these fluctuations diminish depends on the disk viscosity: it is a few years for $\alpha = 10^{-2}$, a few hundred years for constant $\nu = 10^{-5}$, and greater than the simulation time of 500 years for $\alpha = 10^{-3}$. 

For our subsequent simulations we choose $\alpha = 10^{-2}$.  This choice is convenient, as the disk rapidly adjusts to the planet's presence in this case, but is also motivated by observations of T-Tauri disks that are most accurately modeled with $\alpha = 10^{-2}$ \citep{1998ApJ...495..385H, 1998ApJ...507..361S, 2001ApJ...554..391H, 2003A&A...400..185L, 2007ApJ...659..705A}.  This choice of $\alpha$ results in fast migration; i.e., migration on a thousand-year timescale.

\subsubsection{Disk Boundary Conditions}

\begin{figure*}
\includegraphics[width=\textwidth]{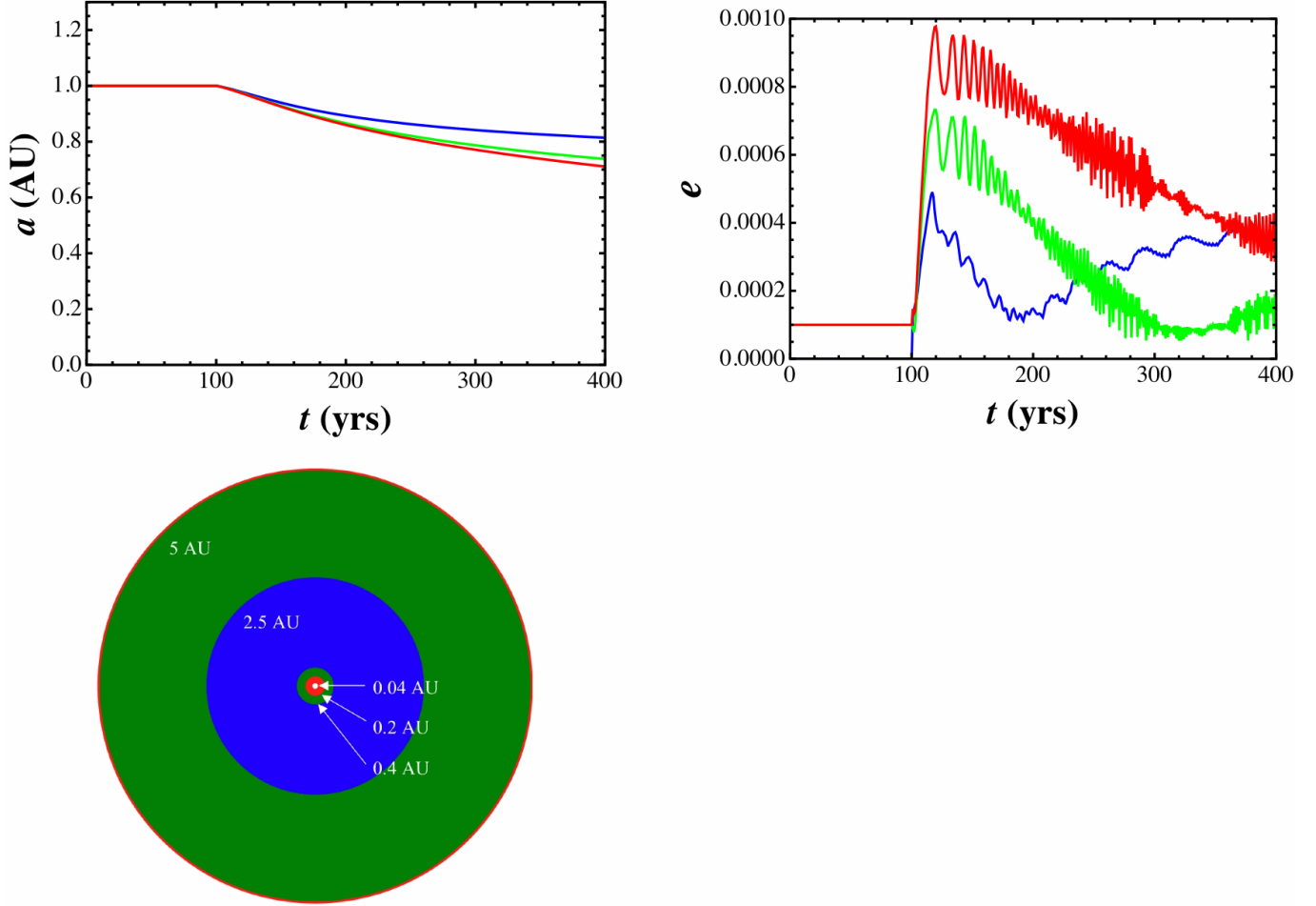}

\caption[Grid size]{Semi-major axis and eccentricity evolution for a planet on an initially circular orbit at 1 AU with varying grid size, color coded to correspond with the diagram at bottom left.}
  \label{gridsize}
\end{figure*}

Next, we present the results of varying the boundary conditions between runs, while keeping the resolution constant.  Extending boundary conditions from [0.4, 2.5] to [0.2, 5.0] to [0.04, 5.0] AU has a small effect on the values of the final semi-major axis and eccentricity of a planet starting in a near-circular orbit (Fig.~\ref{gridsize}) but the overall orbital evolution does not qualitatively change.  (Note that to maintain the same grid resolution, we increase the number of radial grid cells from 128 to 256 to 384.)  

\begin{figure*}
\includegraphics[width=\textwidth]{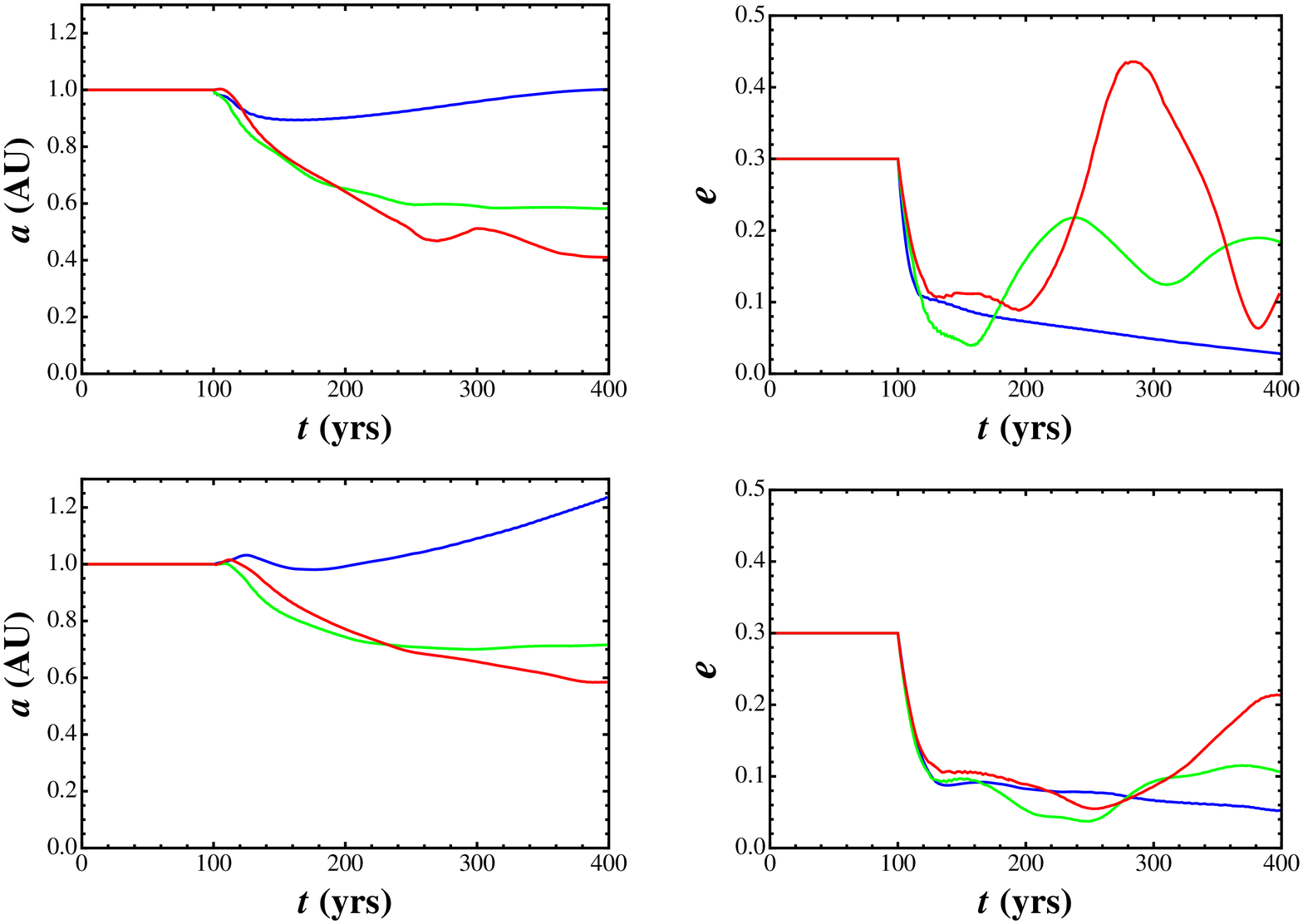}

\caption[Grid size -- eccentric]{Semi-major axis and eccentricity evolution for a planet on an initially eccentric ($e_0 = 0.3$) orbit at 1 AU with varying grid size, color coded to correspond with the diagram in Fig.~\ref{gridsize}.  Simulations shown in the top two plots have non-reflecting boundary conditions, while simulations in the bottom two plots have open boundary conditions. }
\label{bounds}
\end{figure*}

While a planet with low orbital eccentricity stays near $a = 1$ AU, we would like to study planets with appreciable eccentricity; such a planet would pass through a range of radii of [0.7, 1.3] AU.  Fig.~\ref{bounds} depicts simulations with the same set of boundary conditions as those shown in Fig.~\ref{gridsize}; the sole difference is that the planets of Fig.~\ref{bounds} begin with an eccentricity of 0.3.  We find that when planets possess non-negligible eccentricity, the choice of boundary conditions significantly effects the long-term orbital evolution.  Yet $de/dt$ immediately following release of the planet is independent of grid size, and $a(t)$ is similar for the two larger disk cases, suggesting convergence.

The data presented in Figs.~\ref{torques} through \ref{bounds}  are the result of using FARGO's ``non-reflecting boundary conditions," which removes the reflected wake of the planet at the boundary, but does not allow the outflow of disk material.  We perform analogous simulations using open boundary conditions where mass is allowed to flow off the edges of the grid.  Again, we vary the position of these boundary conditions as depicted in Fig.~\ref{gridsize}, and again, we find that varying the boundary locations affects the long-term orbital evolution, but not the initial $de/dt$, of the planet  (Fig.~\ref{bounds}.)  While these plots inform us as to the effects of varying the boundary conditions in hydrodynamical simulations, we encountered a pronounced slowing-down of FARGO before these effects converged.  With this limitation in mind, we will continue to explore the effects of simulation choices using boundaries located at 0.2 and 5 AU and a radial grid resolution of 256.

Animations of a subset of these simulations are available online.\footnote{http://www.astro.ufl.edu/$\sim$altheam/hydro.html}

\subsubsection{Grid Resolution}

\begin{figure*}
\includegraphics[width=\textwidth]{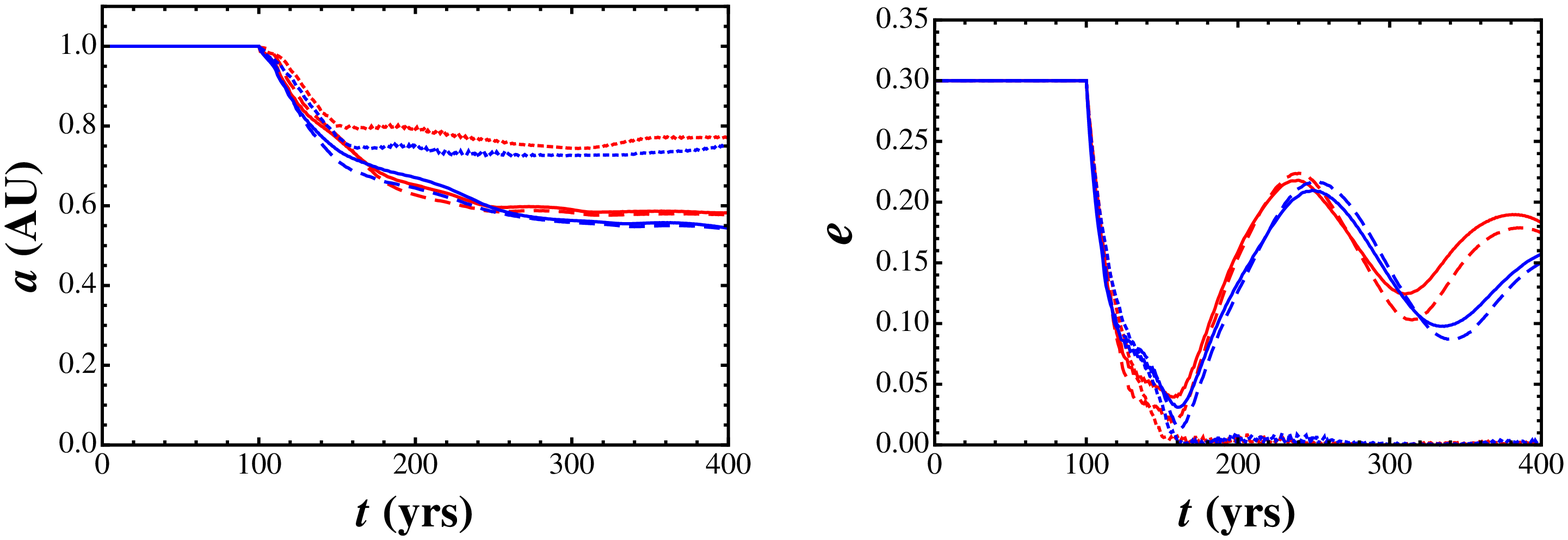}
\caption[Grid resolution]{Orbital evolution for varying azimuthal and radial grid resolution.   The disk boundaries are located at 0.2 and 5.0 AU.  The planet is kept on a fixed orbit for 100 years, at which point the planet begins to migrate under the influence of the disk.  The simulations are coded as follows:
Red curves have a radial grid resolution of 256, blue curves have a radial grid resolution of 128.
 Simulations with an azimuthal grid resolution of 384 are solid, 256 are dashed, and 128 are dotted.
}\label{resolution}
\end{figure*}

In this section we demonstrate that our choice of grid resolution is adequate for calculating eccentricity evolution; care must be taken to ensure that the system is not artificially circularized.  In Fig.~\ref{resolution}, we present a suite of six simulations of a Jupiter-mass planet orbiting for 100 years on a fixed orbit with a semi-major axis of 1 AU and orbital eccentricity of 0.3 which is subsequently released and allowed to migrate under the influence of the disk.   We find that our choice of radial ($N_r = 256$) and azimuthal ($N_{\theta} = 384$) grid resolution is adequate for our simulations; only when the number of azimuthal grid cells is cut to a third of this value do the results vary significantly.  All plots presented from this point forward will depict simulations with $N_r = 256$ and $N_{\theta} = 384$.

\subsubsection{Surface Density}

\begin{figure*}
\includegraphics[width=\textwidth]{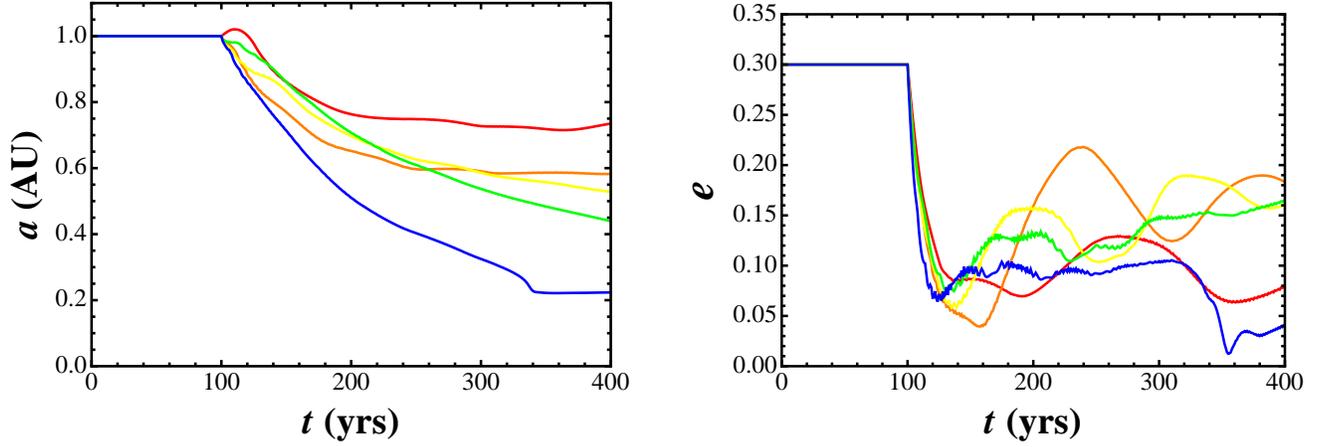}
\caption[Surface density profile]{Orbital evolution for varying $p$, where $\sigma = \sigma_0 (r/r_0)^{-p}$.   The disk boundaries are located at 0.2 and 5.0 AU.  The planet is kept on a fixed orbit for 100 years, at which point the planet begins to migrate under the influence of the disk.  The curves 
are coded as follows: red: $p = -1/2$, orange: $p = 0$, yellow: $p = 1/2$, green: $p = 1$, blue: $p = 3/2$.
 } \label{ps}
\end{figure*}

We found that planet-disk interactions are directly proportional to the surface density, $\sigma_0$ (where the surface density is given by $\sigma = \sigma_0 ~(r/r_0)^{-p}$), yet varying the slope of the surface density profile affects how long the torque oscillations take to settle down.  The steeper the slope, the more persistent these oscillations are.  This, in turn, alters the long-term orbital evolution of the planet's orbit, as can be seen in Fig.~\ref{ps}.  For most disk profiles, the planet drifts inward upon release; however, in one case (p = $-1/2$) the planet initially experiences outward migration.  In all other simulations we adopt $p = 0$ for consistency with other hydrodynamical studies, including that of \cite{2006MNRAS.370..529D}.

\subsubsection{Time of Release}

\begin{figure*}
\includegraphics[width=\textwidth]{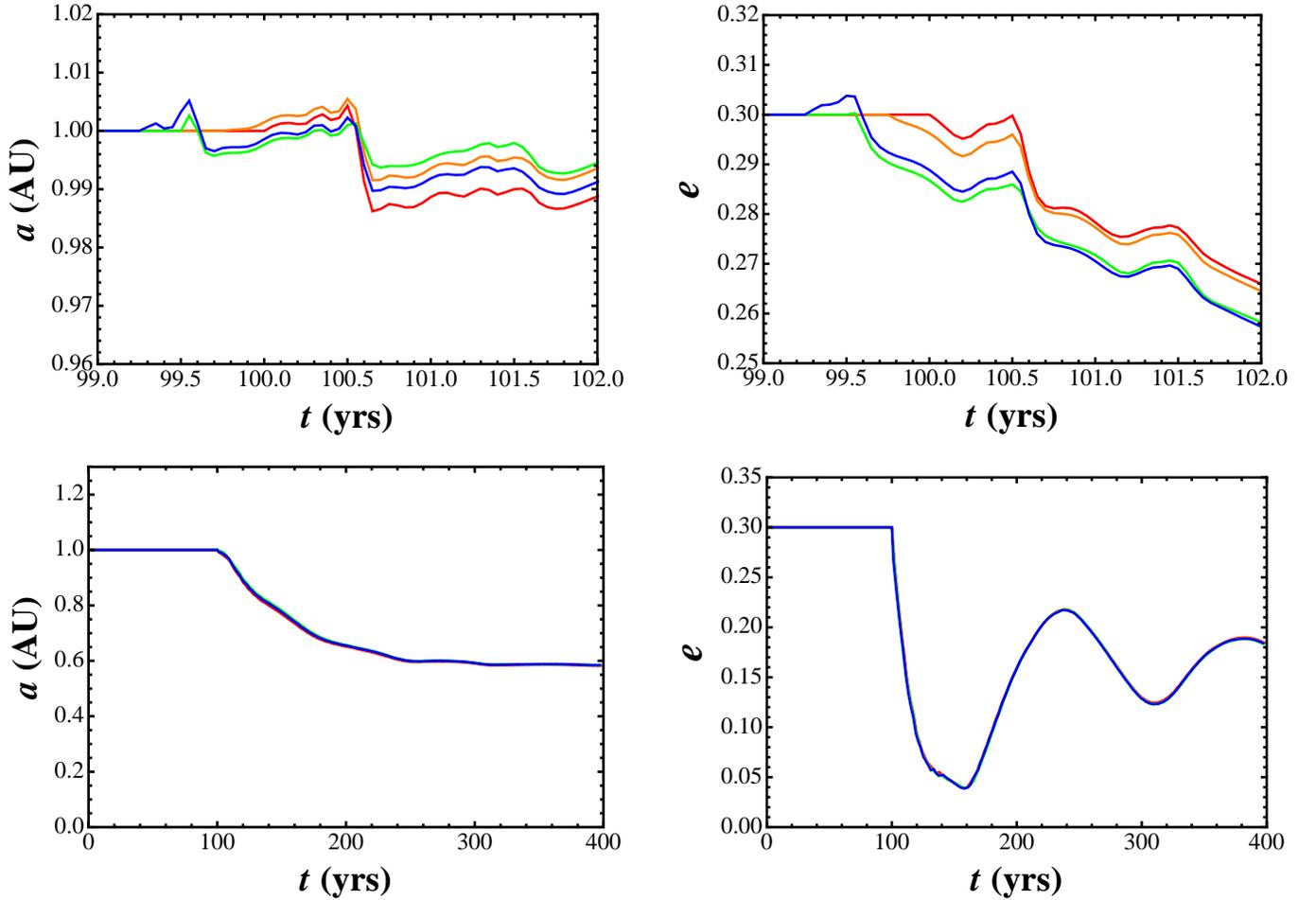}
\caption[Release time]{Semi-major axis (left) and eccentricity (right) evolution of a Jupiter mass planet which is released at different phases in its orbit.  Plots on the top show orbital evolution near the time of release, while plots on the bottom show orbital evolution for the full span of the simulations.   Until 99.3 years have elapsed, the four runs are identical; the runs feature planets released from a fixed orbit at times separated by one quarter of an orbital period.

\vspace{0.5in}}
\label{release}
\end{figure*}

In previous sections we have reported results for simulations in which we released the planet from a fixed, eccentric orbit at the time of apocenter.  Here, we demonstrate that varying the mean anomaly at time of release does not affect the measured $de/dt$ or $da/dt$.  Fig.~\ref{release} shows the eccentricity as a function of time near the time of release for each of these runs.  The time of release in each of these runs differs by one quarter of an orbital period and this difference produces small differences in $e(t)$ and $a(t)$.  However, these differences are indeed small; the right panels of Fig.~\ref{release} show that over tens or hundreds of orbits, the differences in orbital evolution resulting from varying the release time are indistinguishable.  

Additionally, we can see from the left panels of Fig.~\ref{release} that while the semi-major axis and eccentricity appear to be smoothly damped in the lower two panels of Fig.~\ref{release}, this is in fact not the case on sub-orbital timescales.  When we calculate eccentricity and semi-major axis damping (or excitation) timescales for our simulations, we refer to the orbital evolution averaged over several orbits.

\subsection{Orbital evolution of an eccentric planet}
\label{sec:orbit}

Our primary goal in this study is to determine the orbital evolution of an eccentric planet in a thin, flat disk.  In the previous section, we presented results of simulations with a variety of numerical parameters and disk properties, focusing on the semi-major axis and eccentricity evolution.  We found that eccentricity damping occurred upon release of the planet regardless of resolution, disk size, time of release, and surface density profile. We also note that the initial value of $de/dt$ appears to be relatively independent of disk properties and numerical parameters, and thus we may draw conclusions that are insensitive to these quantities.  However, changing the disk's properties -- in particular, the location and type of boundaries -- can substantially alter the long term orbital evolution of an eccentric planet. 

\begin{figure}
\includegraphics[width=\columnwidth]{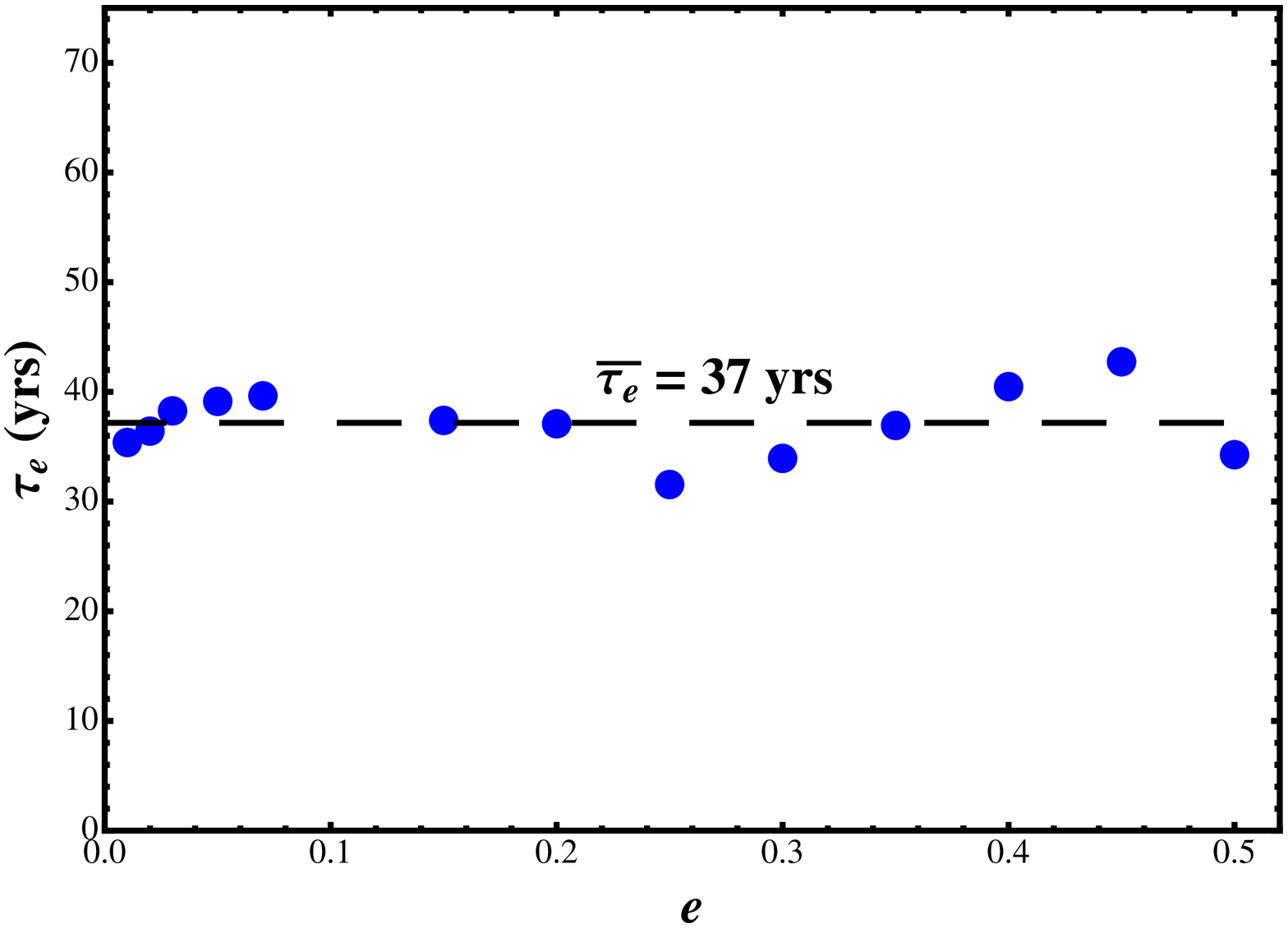}
\caption[$de/dt$ as a function of $e$]{Eccentricity damping timescale as a function of starting eccentricity for a set of hydrodynamical simulations (dots). The average value of this timescale for all shown simulations is underlaid as a dashed line. }
\label{edots}
\end{figure}

In this section we calculate the initial damping rate ($de/dt$) as a function of initial eccentricity for an eccentric planet-disk system.  Previous plots show one of two cases, $e_0 = 0$ or $e_0 = 0.3$.  In Fig.~\ref{edots}, we present $de/dt$ as a function of $e_0$.  The simulations depicted have the following properties: a constant alpha-viscosity of $\alpha = 10^{-2}$, a flat surface density profile ($p = 0$), non-reflecting boundaries at 0.2 and 5.0 AU, and 256 radial and 384 azimuthal grid divisions.  We find that eccentricity is damped on a fairly uniform timescale, $\tau_e \approx 40$ years.  This is longer than an orbital timescale, but shorter than the migration timescale $\tau_a$, justifying our approach of holding the planet on a fixed orbit for 100 years while the disk relaxes.

The eccentricity damping timescale ($\tau_e \approx 40$ years) is independent of initial eccentricity and roughly 15 times faster than the migration timescale of a planet in a circular orbit; the instantaneous value of $(de/dt) / (da/dt)$ upon release varies widely due to varying initial values of $da/dt$; some simulations produce an initial outward drift before inward migration takes place (see, for instance, the $p = -1/2$ case in Fig.~\ref{ps}.)

\section{Conclusions}
\label{sec:conclude}

This study considered hydrodynamic simulations of Jupiter-mass planets in a disk with varying initial orbital eccentricity.  In our simulations, the planet was initially held on a fixed orbit for a period of 100 years, allowing a gap to form around the planet's orbit before measuring $da/dt$ and $de/dt$. Such a scenario could arise for a single planet in a disk due to the following processes: [1] The planet my form via gravitational instability in an eccentric orbit \citep{2000ApJ...536L.101B}. [2] The planet may be scattered by another body before the disk dissipates and forming a gap in the disk before its eccentricity is completely damped \citep{2008ApJ...686..580C, 2008ApJ...688.1361M, 2008ApJ...675.1538T, 2009arXiv0903.2660M}.  [3] The planet may enter a mean motion resonance with another planet; pairs of planets migrating inward while locked in a mean motion resonance experience an increase in orbital eccentricity until an equilibrium is reached between the eccentricity pumping of the migrating resonance and the eccentricity damping of the disk \citep{2002ApJ...567..596L, 2005A&A...437..727K}.

We find that the choice of disk parameters can strongly affect the long-term (i.e., $\gtrsim 10$ yrs) orbital evolution of the planet.  For instance, low disk viscosity produces long lived oscillations in the disk torque.  The placement of the boundaries affects whether and when a resurgence in orbital eccentricity occurs, although this resurgence is less pronounced for open boundary conditions than it is for non-reflecting boundary conditions.  If a very low grid resolution is used, the planet's orbit is artificially circularized, but this is not a concern for the grid resolution of $N_r = 256$, $N_{\theta} = 384$ used for this study.  Finally, we find that the shape  of the surface density profile affects the amplitude and period of oscillations in eccentricity after release, as well as the extent and timescale of orbital migration.

Despite the variety in the orbital evolution of the planet, we find that the initial drop in orbital eccentricity (on timescales of a few to tens of orbits) is consistent across the tested disk parameters.  Furthermore, the eccentricity decay timescale $\tau_e = -e/(de/dt)$ at the time of the planet's release is equal to 40 years and is independent of starting eccentricity for eccentricities up to 0.5 (Fig.~\ref{edots}).  Simulating planets with eccentricities greater than 0.5 was not possible with our chosen disk size.  This value of 40 years is similar to the timescale for eccentricity damping in Type I migration obtained by \cite{2007A&A...473..329C}.  We measured $\tau_a / \tau_e$ (where $\tau_a = -a/(da/dt)$) to be $\lesssim 15$.  While eccentricity damping took place on a shorter timescale than orbital migration, this falls short of the value of 40 required to reproduce the GJ 876 system \citep{2002ApJ...567..596L, 2005A&A...437..727K}.  This scenario of rapid eccentricity damping and less rapid inward migration is not dissimilar to Type I migration, and thus suggests no reason for a difference in orbital element distribution between gas giant and rocky planets. 

Most hydrodynamical treatments of planet-disk interactions during Type I and Type II migration result in a net decrease in eccentricity, and no hydrodynamical treatment increases the planet's orbital eccentricity beyond 0.15 \citep[see, for example,][]{2006ApJ...652.1698D, 2007A&A...473..329C}.  Thus, we conclude that disk-planet interactions alone are unlikely to excite the eccentricities of exoplanets and to produce the observed range in eccentricity for giant planets. 

Planet disk interactions may be important for damping eccentricities during the formation of multiple-planet systems; eccentricity damping is often required to reproduce the observed range in orbital elements \citep{2005Icar..178..517M, 2008ApJ...686..580C, 2008ApJ...675.1538T, 2009arXiv0903.2660M}.  Furthermore, planet-disk interactions may be important for setting the final orbital configuration of resonant planetary systems \citep{2002ApJ...567..596L, 2005A&A...437..727K, 2006A&A...451L..31S, 2007A&A...472..981S}.  Future research should consider a combination of planet-planet interactions and planet-disk interactions, using our prediction damping timescales of  $\tau_a / \tau_e = 15$, to determine if these processes can reproduce the  observed distribution in exoplanet orbital elements.  Additionally, we will investigate the degree to which self-gravity affects planet-disk interactions in the eccentric regime, using a recently available version of FARGO that includes disk self-gravity \citep{2008ApJ...678..483B}.

\bigskip

\section*{Acknowledgments}

First and foremost, we would like to thank Frederic Masset for making the 2D hydrodynamic grid code FARGO and its support materials publicly available.  We would also like to thank Fred Adams, Frederic Masset, and Wilhelm Kley for beneficial discussions and suggestions regarding this work.

We would like to acknowledge the University of Florida High-Performance Computing Center\footnote{http://hpc.ufl.edu} for providing computational resources and support that have contributed to the research results reported within this paper.

This research was supported by the University of Florida and NASA JPL subcontract number 1349281.

\bibliography{adssample}

\end{document}